\documentclass[%
reprint,
footinbib,
aps,
prb,
]{revtex4-1}

\usepackage{hyperref}
\hypersetup{colorlinks=true, linkcolor=blue, urlcolor=blue, citecolor=blue}
\usepackage{graphicx}
\usepackage{amsmath} \usepackage{amssymb}
\usepackage{accents} \usepackage{mathrsfs} \usepackage{setspace}
\usepackage{color}
\definecolor{dgreen}{RGB}{00, 120, 00} \definecolor{dblue}{RGB}{00, 00, 220}
\definecolor{lgreen}{RGB}{46, 139, 87} 
\usepackage{ulem}

\newcommand{\bs}[1]{\boldsymbol{#1}}

\newcommand{\ut}[1]{\undertilde{#1}} 
 \newcommand{\vphi}[0]{\varphi}

\begin{document}

%
%


\title{Robustness of chiral surface current and subdominant $s$-wave Cooper
pairs}

\author{Shu-Ichiro Suzuki}

\author{Alexander A. Golubov}
\affiliation{MESA+ Institute for Nanotechnology, University of Twente, 
7500 AE Enschede, The Netherlands}

\date{\today}

\begin{abstract}
	The robustness of the chiral surface current of chiral superconductors against surface roughness is studied utilizing the quasiclassical Eilenberger theory. We consider the general chiral superconductors where the pair potential is given by the spherical harmonics $Y_l^m$ such that $(l,m)=(1, \pm1)$ state corresponds to an ($p_x \pm ip_y$)-wave superconductor.  The self-consistent calculations demonstrate that the robustness of the chiral current is determined by whether subdominant $s$-wave Cooper pairs are induced by disorder. The induced $s$-wave pairs act as an effective pair potential. As a result, the spontaneous chiral current of ($p_x+ip_y$)- and ($d_{x^2-y^2}+id_{xy}$)-wave superconductors are robust against the roughness because the subdominant $s$-wave Cooper pairs are present. 
\end{abstract}

\pacs{pacs}

\thispagestyle{empty}

\maketitle

\section{Introduction}
Chiral superconductivity is realised by Cooper pairs with finite orbital angular momenta. The non-zero angular momenta give rise to the so-called chiral current at a surface of a chiral superconductor (SC) spontaneously \cite{ Kita_JPSJ_1998, Matsumoto_JPSJ_1999, Furusaki_PRB_2001, Stone_04, Nagato_11, Sauls_PRB_2011, Bakurskiy_14, Sigrist_PRB_2014, Lederer_PRB_14, Bakurskiy_17, Braunecker_PRL_2005, Huang_PRB_2014, Tada_PRL_2015, suzuki_16, Wang_PRB_2018, Higashitani_JPSJ_20, Nie_PRB_20, SIS_PRR_2022}. The direction of the chiral current is chosen by the Cooper-pair condensate with the spontaneous symmetry breaking.  Observing the spontaneous surface current can be conclusive evidence to demonstrate the realisation of chiral superconductivity.  However, the spontaneous current has never been observed in any chiral SCs discovered so far \cite{moler_05, nelson_07}. 

The simplest chiral superconducting state is the $p_x+ip_y$-wave state
(e.g., $^3$He-A). The intrinsic total angular momentum in the
$p_x+ip_y$-wave state has been a longstanding problem. In addition to
the $p_x+ip_y$-wave state, other types of chiral SC have been so far
proposed as a paring symmetry in several unconventional
superconductors \cite{Braunecker_PRL_2005, Huang_PRB_2014,
Tada_PRL_2015, suzuki_16, Wang_PRB_2018, Higashitani_JPSJ_20,
Nie_PRB_20, SIS_PRR_2022, kobayashi_15, tamura_17, suzuki_20}. The
possibility of the two-component $d$-wave superconductivity (e.g.,
$d_{x^2-y^2}+id_{xy}$-wave pairing) has been discussed in layered
materials such as Na$_x$CoO$_2$$\cdot y$H$_2$O \cite{Tanaka_PRL_2003,
Baskaran_PRL_2003, takada2003superconductivity, Ogata_JPSJ_2003},
doped graphene \cite{Kiesel_PRB_12, BS_PRL_12, Nand_NatPhys_12,
BS_JPhys_14}, and SrPtAs \cite{ Biswas_PRB_13, Fischer_PRB_14,
Landaeta_PRB_16, Goryo_PRB_17, Ueki_PRB_19}.  In addition,
three-dimensional pairings for the two-component $d$-wave
superconductivity are possible. The $d_{zx}+id_{yz}$-wave pairing has
been discussed to reveal the superconductivity in a uranium compound
URu$_2$Si$_2$ \cite{Kasahara_PRL_07, Kasahara_NJP_09,
Kittaka_JPSJ_16}.  This pairing has recently been proposed as a new
candidate for an unconventional SC Sr$_2$RuO$_4$ \cite{maeno_94,
maeno_03, Pustogow_Nature_19, Agterberg_PRR_2020, Grinenko_20,
Grinenko_21, Ikegaya_PRR_21, SIS_PRR_2022, Yuri_PRR_2022}.
Furthermore, the possibility of the spin-triplet $f+if'$-wave
superconductivity has been discussed to elucidate the
superconductivity in UPt$_3$
\cite{Fisher_PRL_1989, Sauls_PRL_1991, Sauls_JLTP_1994,
Machida_JPSJ_1999, Sauls_PRB_2000, Joynt_RMP_02, Machida_PRL_12,
Tsutsumi_JPSJ_12, Izawa_JPSJ_14, Lambert}.

To achieve the observation of the spontaneous chiral current, we
cannot avoid the effects of surface roughness. In real-life
experiments, the sample quality at a surface is not as specular as
assumed in theoretical models. Although the total chiral current is
estimated by assuming the secular surface \cite{ Matsumoto_JPSJ_1999,
Furusaki_PRB_2001}, the surface roughness causes random reflections
that is known to affect the surface Andreev bound states \cite{ Nagato_11,
Bakurskiy_14, suzuki_16, SIS_PRR_2022, Yamada_JPSJ_96, Nagai_JPSJ_08,
SIS_PRB_15, Higashitani_JPSJ_15} in unconventional SCs such as
high-$T_c$ SCs. 
Indeed, it has been shown that the surface quality significantly
modifies the amount of the chiral current for higher-order chiral SCs
\cite{suzuki_16}. Moreover, the suppression of the chiral current is
deeply related to the pairing symmetry of the chiral SC. In the
previous paper \cite{SIS_PRR_2022}, the robustness of the chiral
surface currents in the $p_x+ip_y$-wave and $d_{zx}+id_{yz}$-wave SCs
has been compared.  It has been demonstrated that the chiral current
for $p_x+ip_y$-wave SC can survive under the surface roughness,
whereas that for the $d_{zx}+id_{yz}$-wave SC disappears even under a
weak roughness. The difference is well explained by the existence of
the subdominant $s$-wave pairs at the surface. The $s$-wave pairs
appear at a surface of the $p_x+ip_y$-wave SC and act as an effective
pair potential, leading to the robust chiral surface current of the 
$p_x+ip_y$-wave SC. In the
$d_{zx}+id_{yz}$-wave case, on the contrary, no $s$-wave pair is
induced, resulting in the fragile surface chiral current. 

\begin{figure}[b]
	\includegraphics[width=0.46\textwidth]{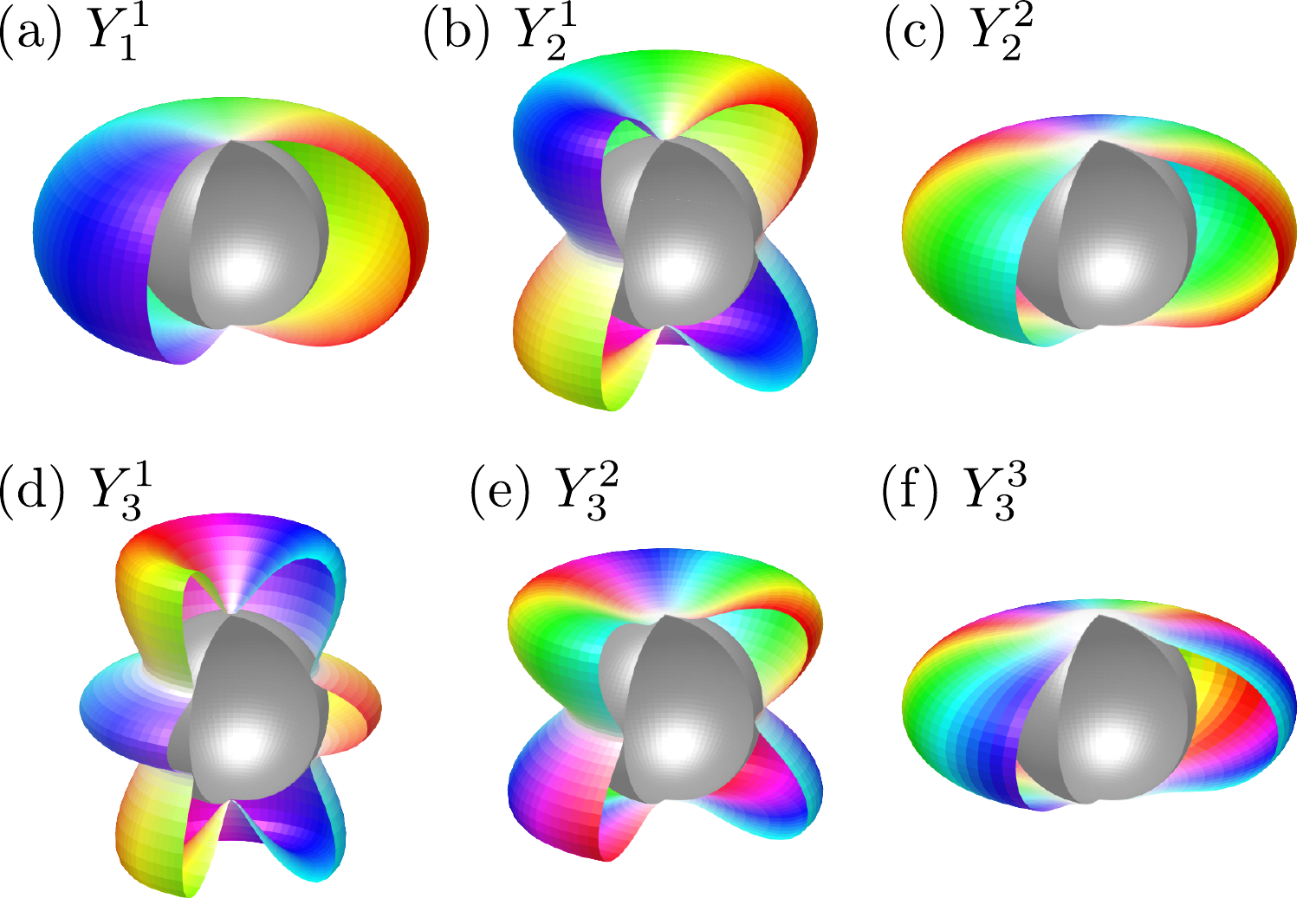}
	\caption{Schematic superconducting gaps in the homogeneous limit.
	  The superconducting gap of a general chiral superconductor is
		given by the spherical harmonics $Y_l^m$ with $m>0$. 
		The color means the phase of the pair potential. The inner sphere
		indicates the Fermi sphere. }
  \label{fig:gap}
\end{figure}

In this paper, we examine the robustness of the spontaneous surface
current in general chiral SCs, in particular,  by focusing on the
subdominant $s$-wave pairs at the surface.  The pair potential is
given in a general form using the spherical harmonics $Y_l^m$ ($m \geq
1$) as shown in Fig.~\ref{fig:gap}. Chiral SCs with rough surfaces and
that covered with dirty normal metals are considered
(Fig.~\ref{fig:Sche}). Solving the Eilenberger equation with the
self-energy by random scatterings at the surface, the spatial profiles
of the pairs potentials, current density, and the $s$-wave pair
amplitude are determined self-consistently. 

\begin{table*}[tb]
	\caption{Momentum dependence of the pair potential and attractive
	potentials.  We use the modified spherical harmonics so that
	$\textrm{max}[Y_l^m(\bs{k})]=1$. The amplitude of each component
	$\Delta_{1(2)}$ is determined by the self-consistent gap equation. }
  \label{t:table}
  \centering
	\includegraphics[width=0.96\textwidth]{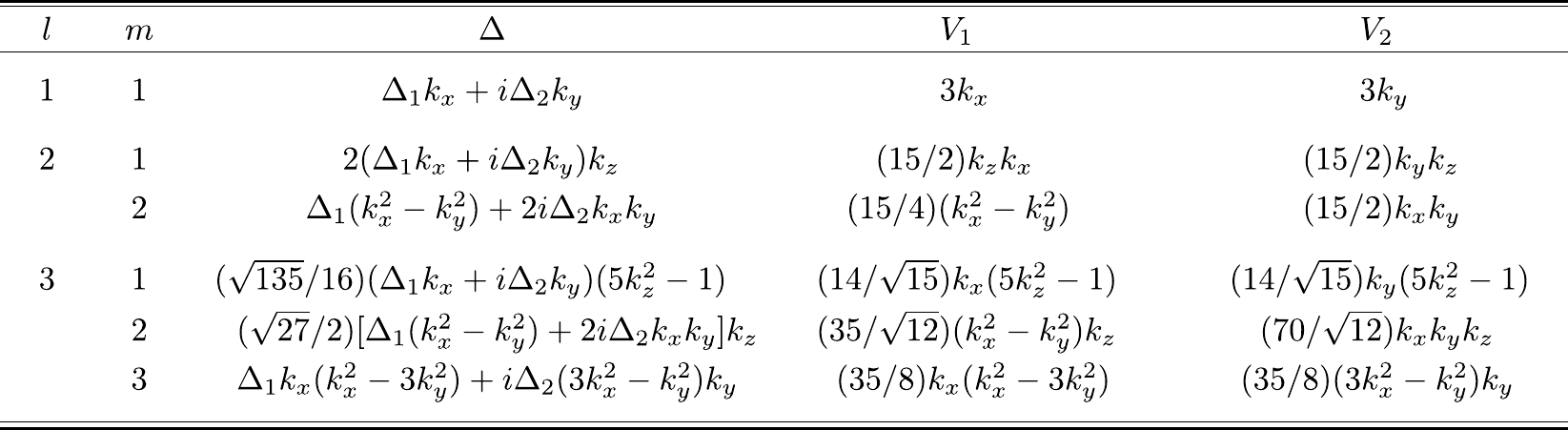}
\end{table*}

We conclude that the chiral surface current in the $Y_1^1$, and $Y_2^2$, and $Y_3^3$ [i.e., $p_x+ip_y$-, $d_{x^2-y^2}+id_{xy}$-, and $f_{x(x^2-3y^2)} + i f_{y(3x^2-y^2)}$-wave] SCs can survives under the strong-roughness limit. In particular, the chiral currents in $Y_1^1$ and $Y_2^2$ SCs are sufficiently large to be observed in experiments.  The chiral current in $Y_3^3$ SC remains finite but could not be large enough to be detected.  In the other chiral SCs, specular surfaces are required to observe the spontaneous current because the surface currents in these cases disappear even with a weak roughness. 
We also confirm the subdominant $s$-wave Cooper pairs are essential for the observable spontaneous surface current by comparing the calculated current density with and without the subdominant $s$-wave Cooper pairs.  We show that, if there were no $s$-wave Cooper pairs, the spontaneous chiral currents for all chiral SCs can not survive even under weak surface roughness.  Namely, the robustness of the spontaneous current can be judged only by whether the subdominant $s$-wave Cooper pairs emerge.

\section{Quasiclassical Eilenberger theory}
\begin{figure}[tb]
	\includegraphics[width=0.48\textwidth]{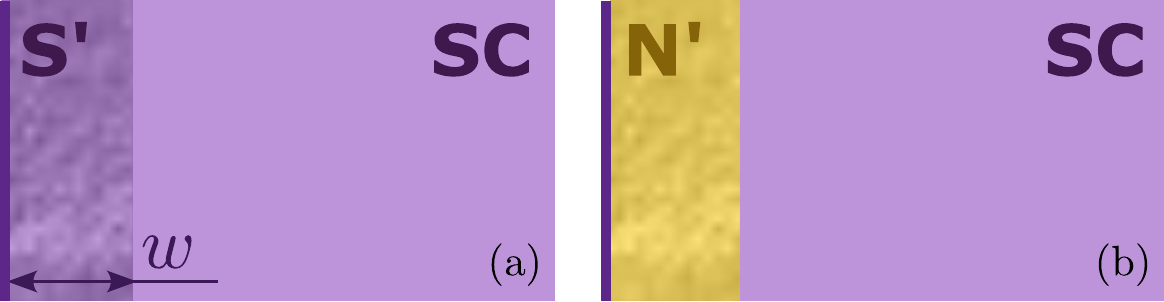}
	\caption{Schematics of the systems. Superconductors with (a) rough
	surfaces and (b) dirty metallic surfaces. The disordered superconducting
	and normal-metal regions are indicated by S' and N'. 
	The surfaces are perpendicular to
	the $x$ axis and located at $x=0$ and $2L$. The width of the
	disordered region is denoted by $w$. 
	The system in (b) corresponds to the so-called Ovchinnikov model. 
	}
  \label{fig:Sche}
\end{figure}

We examine the effects of surface roughness utilizing the
quasiclassical Eilenberger theory \cite{Eilenberger} in equilibrium. 
\footnote{The quasiclassical approximation allows us to extract the
essential spatial profile (i.e., coherence-length order) from the
Green's function by ignoring the rapid oscillation with the
Fermi-wavelength order.} 
The SC has a pair of parallel surfaces which are perpendicular to the
$x$ axis. The distance between two surfaces is denoted by $L$.  The
thin dirty regions and thin dirty normal metals with the width $w$ are
introduced at the surfaces as shown respectively in Fig.~\ref{fig:Sche}(a) and
\ref{fig:Sche}(b). 
The superconductor covered by a thin dirty normal metal corresponds to
the so-called Ovchinnikov model \cite{Ovchinnikov, Golubov_Ovchi98, Golubov_Ovchi99}. 
The Green's functions obey the
Eilenberger equation which is valid in the weak-coupling limit, 
\begin{align}
  & i \bs{v}_F \cdot \bs{\nabla} \check{g}
	+ \left[ \, i\omega_n \check{\tau}_3+\check{H},~\check{g} \right]_-
	= 0, 
	\label{eq:Eilen}
	\\[2mm]
	& \check {g} 
	= \left( \begin{array}{rr}
	 \hat{    g}  &  
	 \hat{    f}  \\[1mm]
	-\hat{\ut{f}} & 
	-\hat{\ut{g}} \\
	\end{array} \right), 
	\hspace{4mm}
	 \check {\Delta} = \left( \begin{array}{cc}
	 0&  
	 \hat{\Delta} \\[1mm]
	 \hat{\ut{\Delta}} & 
	 0\\
	\end{array} \right), 
	\\[2mm]
  & \check {H}
  =\check {\Delta}+ \check {\Sigma}
	= \left( \begin{array}{rr}
	  \hat{\xi } &  
	  \hat{\eta} \\[1mm]
	  \hat{\ut{\eta}} & 
	  \hat{\ut{\xi }} \\
	\end{array} \right), 
	\hspace{4mm}
  \check {\Sigma}
	=  \frac{1}{2 \tau_0}
	 \langle \check{g}\rangle , 
\end{align}
where $
  \langle \cdots \rangle
  = \int_0^\pi \int_{-\pi}^\pi \cdots 
	{\sin \theta d\vphi d\theta}/{4 \pi} $ 
	is the angle average on the Fermi sphere, the unit vector $\bs{k}=(
\sin \theta \cos \phi, 
\sin \theta \sin \phi, 
\cos \theta
)$ represents the direction of the Fermi momentum, 
$\check{g}= \check{g} (\bs{r},\bs{k},i\omega_n)$ is the quasiclassical
Green's function in the Matsubara representation, 
$\check {\Delta} = \check {\Delta}(\bs{r},\bs{k})$ is the pair-potential matrix, 
$\check {\Sigma} = \check {\Sigma}(\bs{r},i \omega_n)$ is the
self-energies with the mean free path $\ell = v_F \tau_0$.  In this paper, the
accents $\check{\cdot}$ and $\hat{\cdot}$ means matrices in
particle-hole and spin space.  The identity matrices in particle-hole
and spin space are respectively denoted by $\check{\tau}_0$ and
$\hat{\sigma}_0$.  The Pauli matrices are denoted by $\check{\tau}_\nu$
and $\hat{\sigma}_\nu$ with $\nu \in \{1, 2, 3 \}$. 
All of the functions satisfies the symmetry relation $\hat{K}
(\bs{r},\bs{k},i \omega_n) = [\hat{\ut{K}} (\bs{r},-\bs{k},i
\omega_n)]^*$. Effects of the vector potential are ignored
because it affects on surface states only quantitatively. 
The quasiclassical Green's function is supplemented by the
normalization condition 
\begin{align}
  \check {g} 
  \check {g} 
	=
	\check{\tau}_0. 
\label{}
\end{align}
Throughout this paper, we use the units $k_B=\hbar=1$. 
%
%

The Eilenberger equation \eqref{eq:Eilen} can be simplified by the
so-called Riccati parameterization \cite{Schopohl_PRB_95, 
Eschrig_PRB_00, Eschrig_PRB_09}. The Green's function can be
expressed in terms of the coherence function $\hat{\gamma}=
\hat{\gamma}(\bs{r}, \bs{k}, i \omega_n)$, 
\begin{align}
	& \check {g} = 2 \left( \begin{array}{rr}
	 \hat{\mathcal{G}} &  
	 \hat{\mathcal{F}} \\[1mm]
	-\hat{\ut{\mathcal{F}}} & 
	-\hat{\ut{\mathcal{G}}} \\
	\end{array} \right) - \check{\tau}_3, 
	\\[1mm]
	& \hat{\mathcal{G}} = (1-\hat{\gamma} \hat{\ut{\gamma}} )^{-1},\hspace{6mm}
	  \hat{\mathcal{F}} = (1-\hat{\gamma} \hat{\ut{\gamma}})^{-1}\hat{\gamma}.
\label{eq:Ric-Para}
\end{align}
The equation for $\hat{\gamma}$ is given by 
\begin{align}
  & (i \bs{v}_F \cdot \bs{\nabla} + 2 i \omega_n ) \hat{\gamma}
	+ \hat{\xi} \hat{\gamma} - \hat{\gamma}\hat{\ut{\xi}}
	-\hat{{\eta}}+ \hat{\gamma} \hat{\ut{\eta}} \hat{\gamma}= 0. 
	\label{eq:Riccati03}
\end{align}
Assuming no spin-dependent potential and single-spin $\hat{\Delta}$, 
we can parameterize the spin structure of the functions, 
\begin{align}
	& \hat{\Delta} = i \Delta_{\bs{k}, \nu} (i \hat{\sigma}_\nu \hat{\sigma}_2), 
	\\
	& \hat{\ut{\Delta}} = -i \Delta_{-\bs{k}, \nu}^* (i \hat{\sigma}_\nu \hat{\sigma}_2)^*
	= i \Delta_{\bs{k}, \nu}^* (i \hat{\sigma}_\nu \hat{\sigma}_2)^\dagger
	\\
	&     \hat{g}  =     g         \hat{\sigma}_0, ~ ~
	      \hat{f}  =     f _\nu (i \hat{\sigma}_\nu \hat{\sigma}_2), ~ ~
		\ut{\hat{f}} = \ut{f}_\nu (i \hat{\sigma}_\nu \hat{\sigma}_2)^\dagger, 
		\\
  & \hat{\eta} = i \eta_{\nu} (i \hat{\sigma}_\nu \hat{\sigma}_2),
	~ ~ 
\ut{\hat{\eta}} = i \ut{\eta}_{\nu} (i \hat{\sigma}_\nu \hat{\sigma}_2)^\dagger, 
\end{align}
where $\nu = 0$ ($\nu \in \{1, 2, 3 \}$) is for the spin-singlet (spin-triplet) SC. In the following, we make $\nu$ explicit only when necessary. 
Equation~\eqref{eq:Riccati03} can be reduced to 
\begin{align}
  & \bs{v}_F \cdot \bs{\nabla}  \gamma
	+ 2 \tilde{\omega} \gamma
	  -     \eta 
		+ \ut{\eta}
	\gamma^2= 0, 
	\label{eq:Riccati07}
	\\
	& \tilde{\omega} = 
	\omega_n 
	+\frac{\mathrm{Re} \langle g \rangle }{2 \tau_0}, 
	\\
	& \eta_{\nu} = \Delta_{\bs{k}} + \frac{\langle f \rangle}{2 \tau_0}, 	
	\hspace{6mm} \ut{\eta}_{\nu} = \Delta_{\bs{k}}^* -S_\nu
	\frac{\langle f \rangle^*}{2 \tau_0}. 
  \label{eq:ep}
\end{align}
with $S_\nu = +1$ ($-1$) for the spin-triplet (spin-singlet) SC. 
In the homogeneous limit, $\gamma$ is given by
\begin{align}
  & 
  \bar{\gamma} (\bs{k}, i \omega_n) = 
	\frac{s_o \Delta_{\bs{k}} }
	{ |\omega_n| + \sqrt{\omega_n^2 + |\Delta_{\bs{k}}|^2} }, 
\end{align}
with $s_o=\mathrm{sgn}[\omega_n]$ and $\bar{\cdot}$ means the bulk
value.

\begin{figure}[tb]
	\includegraphics[width=0.42\textwidth]{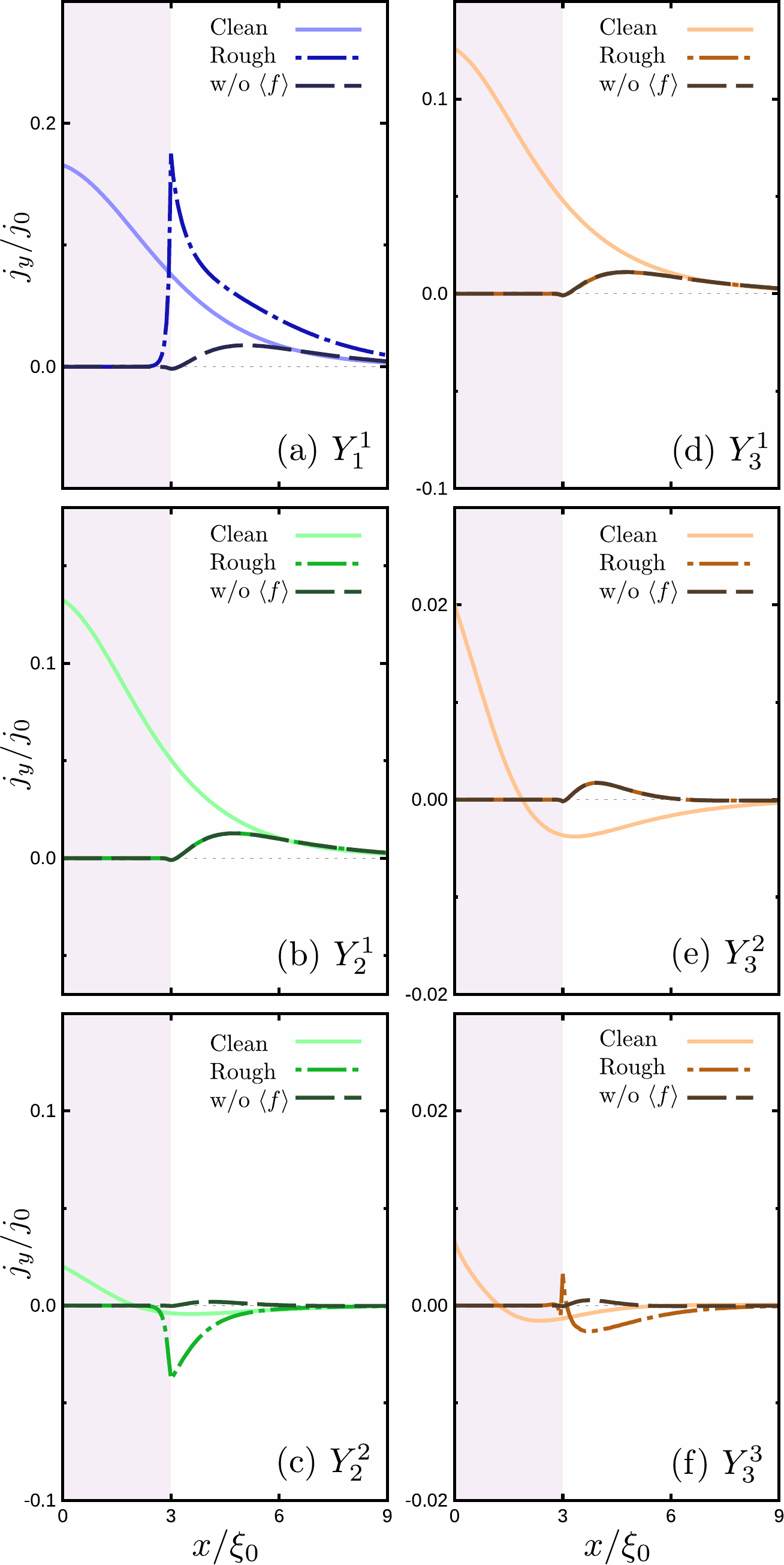}
	\caption{Partial current densities at the lowest Matsubara frequency 
	$j_{y,0}$ in each chiral SC. The results for the clean surface,
	rough surface, and rough surface without $\langle f \rangle$ are
	plotted with the solid (Clean), one-dot chain (Rough), and 
  broken lines (w/o $\langle f \rangle$). The roughness parameters are set to
	$\xi_0/\ell=5$ and $w=3\xi_0$.  The shaded regions indicate the
	surface disordered region. 	
	The pair potential is assumed 
	(a) $Y_1^1$, (b) $Y_2^1$,    ( c) $Y_2^2$, 
	(d) $Y_3^1$, (e) $Y_3^2$, and (f) $Y_3^3$. 
	When there was no $s$-wave Cooper pairs (i.e., $\langle f
	\rangle$=0), the
	chiral current disappears in all chiral SCs. The current densities
	are normalized to $j_0 = |e|v_F \pi N_0/ \beta$. 
  We fix the parameters: 
  $L=80 \xi_0$, $w=3\xi_0$, $\omega_c = 10
  \pi T_c$, and $T=0.4T_c$
  }
  \label{fig:cur}
\end{figure}

The pair potential of chiral superconductors is described by the
two-component pair potential as summarized in Table~\ref{t:table}. 
Deep inside the superconductor, the momentum dependence of the pair
potential is given by 
\begin{align}
  \Delta_{\bs{k}} = \bar{\Delta} \tilde{Y}_l^m(\bs{k})
  \label{}
\end{align}
where we use the modified spherical harmonics $\tilde{Y}_l^m$ which satisfies 
$\mathrm{max}[\tilde{Y}_l^m(\bs{k})] = 1$. 
The chiral superconductivity is described by $m \geq 1$. 
The schematic gap amplitudes in the bulk are shown in
Fig.~\ref{fig:gap}, where the color means the phase of the pair
potential $\mathrm{arg}[\Delta(\bs{k})]$ and the inner silver sphere
means the Fermi sphere. 
The spatial dependence of $\Delta_1(\bs{r})$ and $\Delta_2(\bs{r})$ 
are determined by the
self-consistent gap equation which relates $f$ and $\Delta$: 
\begin{align}
	& \Delta_\mu(\bs{r})
	=
	2 \lambda N_0 \frac{\pi}{i \beta} \sum_{\omega_n}^{\omega_c}
	\langle V_\mu(\bs{k}') f(\bs{r},\bs{k}',i \omega_n) \rangle, 
\end{align}
where $\mu=1$ or $2$, $\beta=1/T$, $T_c$ is the critical temperature, $N_0$ is the density of the states (DOS) in the normal state at the Fermi energy, and $n_c$ is the cutoff
integer which satisfies $
2n_c+1 < \omega_c/\pi T < 2n_c+3
$. 
The attractive potentials $V_1$ and $V_2$ are also 
summarized in Table~\ref{t:table}. 
The coupling constant $\lambda$ is finite in the superconducting
region, 
\begin{align}
	& \lambda 
	= \frac{1}{2 N_0}
	\left[
	\ln\frac{T}{T_c} 
	+ \sum_{n=0}^{n_c}
	\frac{1}{n+1/2}
\right]^{-1}. 
\end{align}
In the normal metal, the coupling constant is set to $\lambda=0$. 

The spontaneous chiral current is 
calculated from the Green's function, 
\begin{align}
	&j_y(\bs{r})
	= \sum_{\omega_n>0}^{\omega_c} j_n
	\\
	&
	j_n=ev_F \frac{4 \pi N_0}{\beta} 
	\langle k_y
	\mathrm{Im}[{g}(\bs{r},\bs{k},i \omega_n)] \rangle, 
\end{align}
with $e <0$ is the charge of a quasiparticle. 

In the numerical simulations, we fix the parameters: 
$L=80 \xi_0$, $w=3\xi_0$, $\omega_c = 10
\pi T_c$, and $T=0.4T_c$ with $\xi_0 = 
v_F / 2 \pi T_c$ being the coherence length.

\section{Rough surface}

\begin{figure}[tb]
	\includegraphics[width=0.46\textwidth]{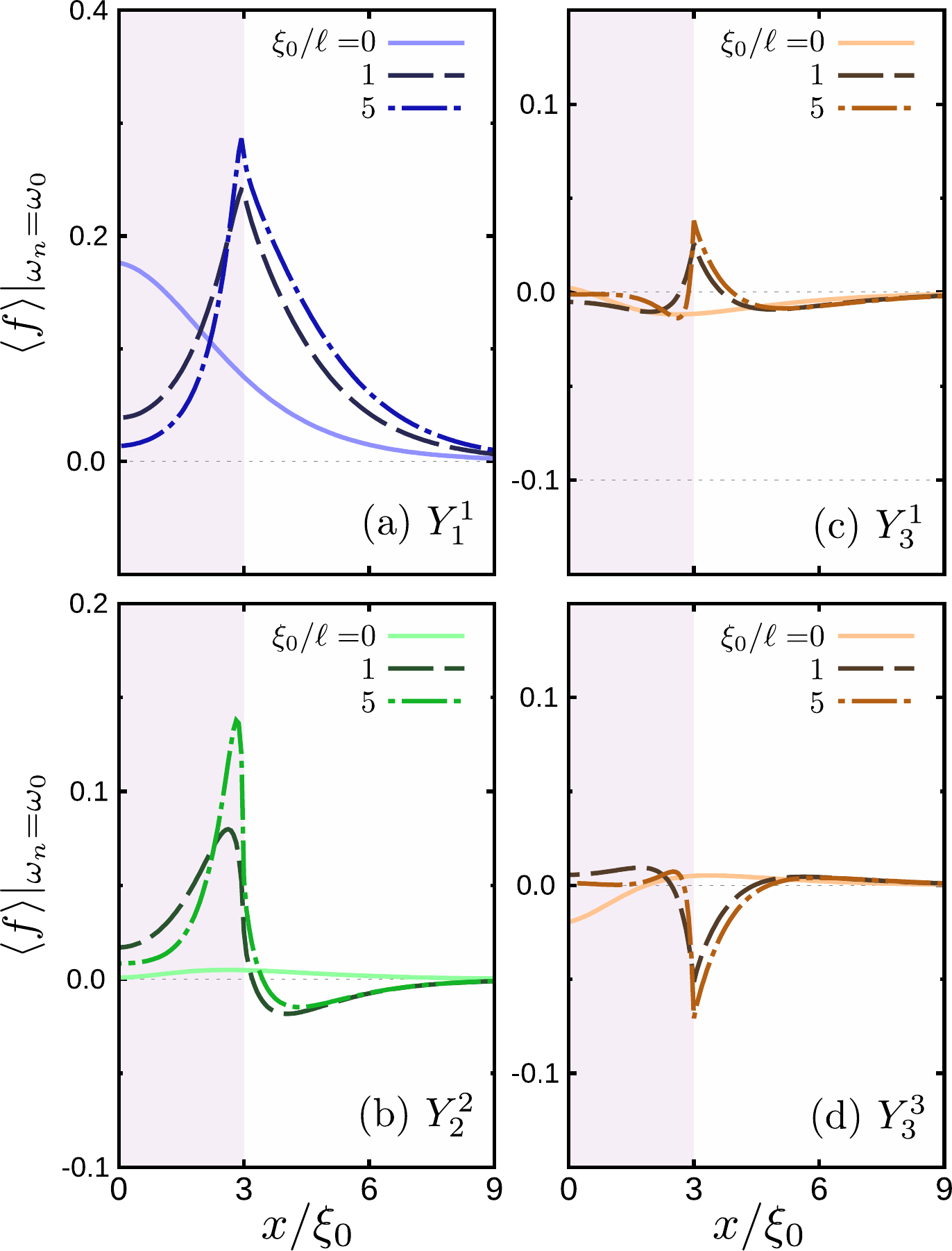}
	\caption{Spatial profiles of the $s$-wave Cooper pairs $\langle f
		\rangle$ at $\omega_0$. 
		The strength of the roughness is set to $\xi_0/\ell = 0$, 1,
		and 5. The $s$-wave pairs have a large
		amplitude in the $Y_1^1$ SC. In (c,d), the $s$-wave pairs have
		finite amplitudes in the disordered region but smaller than those
		in (a,b). 
	  We have confirmed that $\langle f \rangle = 0$ in the
    $Y_2^1$ and $Y_3^2$ SCs because $\Delta_{\bs{k}} \sim k_z$. }
  \label{fig:pair}
\end{figure}

We begin with the chiral surface current with rough surfaces as shown
in Fig.~\ref{fig:Sche}(a). The partial chiral surface currents
$j_{n=0}$ 
for each chiral
SC are shown in Fig.~\ref{fig:cur}. In the clean limit, the chiral
surface currents for the $m=1$ chiral SCs (i.e., $Y_1^1$, $Y_2^1$, and
$Y_3^1$) are sufficiently large to observe in experiments
\cite{Matsumoto_JPSJ_1999, moler_05, nelson_07} as shown in Fig.~\ref{fig:cur}(a,b,d). On the other hand, the chiral
surface currents for the $m>1$ chiral SCs are smaller because the
contributions from each chiral channel compensate each other [see, Figs.~\ref{fig:cur}(c,e,f)]
\cite{Huang_PRB_2014, Tada_PRL_2015, suzuki_16, Wang_PRB_2018,
Higashitani_JPSJ_20, Nie_PRB_20}. 

When the surface is rough, the chiral currents in the $Y_2^1$, $Y_3^1$, and
$Y_3^2$ SCs disappear by the roughness is shown in
Fig.~\ref{fig:cur}(b,d,e)
where the ratio of the coherence length and the mean free path is set
to $\xi_0/\ell=5$. On the other hand, the chiral
currents in the $Y_1^1$, $Y_2^2$, and $Y_3^3$ SCs survive even under
the rough surface [Fig.~\ref{fig:cur}(a,c,f)]. The chiral
surface current in the $Y_3^3$ SC is, however, small both of the clean
and rough surface cases. 
Therefore, in a sample with non-specular surfaces, we can
observe the chiral surface current only when the sample is a $Y_1^1$
SC (i.e., $p_x+ip_y$-wave SC) or $Y_2^2$ SC (i.e.,
$d_{x^2-y^2}+id_{xy}$-wave SC). 

In the previous paper \cite{SIS_PRR_2022}, we pointed out that the
subdominant $s$-wave Cooper pairs $\langle f \rangle$ play an
important role under the disorder. 
The spatial dependences of the $s$-wave Cooper
pairs $\langle f \rangle$ in each chiral SC are shown in
Fig.~\ref{fig:pair}, where $\omega_n=\omega_0$ and the strength of the disorder is set to $\xi_0/\ell
= 1$, $3$, or $5$. In the $Y_2^1$ and $Y_3^2$ SCs (i.e., chiral SCs with
fragile chiral currents), the amplitude of the $s$-wave Cooper pairs
is exactly zero because of the $k_z$-dependence of $\Delta_{\bs{k}}
\sim k_z$ (therefore the results are not shown). When the chiral
current is robust, there are always $s$-wave Cooper pairs induced by
the disorder. In the $Y_1^1$ and $Y_2^2$ cases, the large amplitude of
the $s$-wave pairs $\langle f \rangle$ [Fig.~\ref{fig:pair}(a,b)] results in the robust chiral surface current. 
The $s$-wave Cooper pairs act as an effective pair potential in the
disordered region [see Eq.~\eqref{eq:ep}]. In this case, the internal interface between the clean
and disordered region can be regarded as an interface between
an effective $s$-wave SC and a chiral SCs without a
barrier potential. The quasiparticles feel this effective interface 
and give
rise to the spontaneous current along it. 
In the chiral $f$-wave SCs ($Y_3^1$ and $Y_3^3$), the $s$-wave pairs
in the disordered region are present but the amplitude is much smaller
than the $Y_1^1$ and $Y_2^2$ cases. The $s$-wave pairs have the
largest amplitude inside the clean region ($x>w$). In
this case, the $s$-wave pairs cannot affect the quasiparticle as
indicated by Eq.~\eqref{eq:ep} because $\ell \to \infty$ in the clean
region.

\begin{figure}[tb]
	\includegraphics[width=0.46\textwidth]{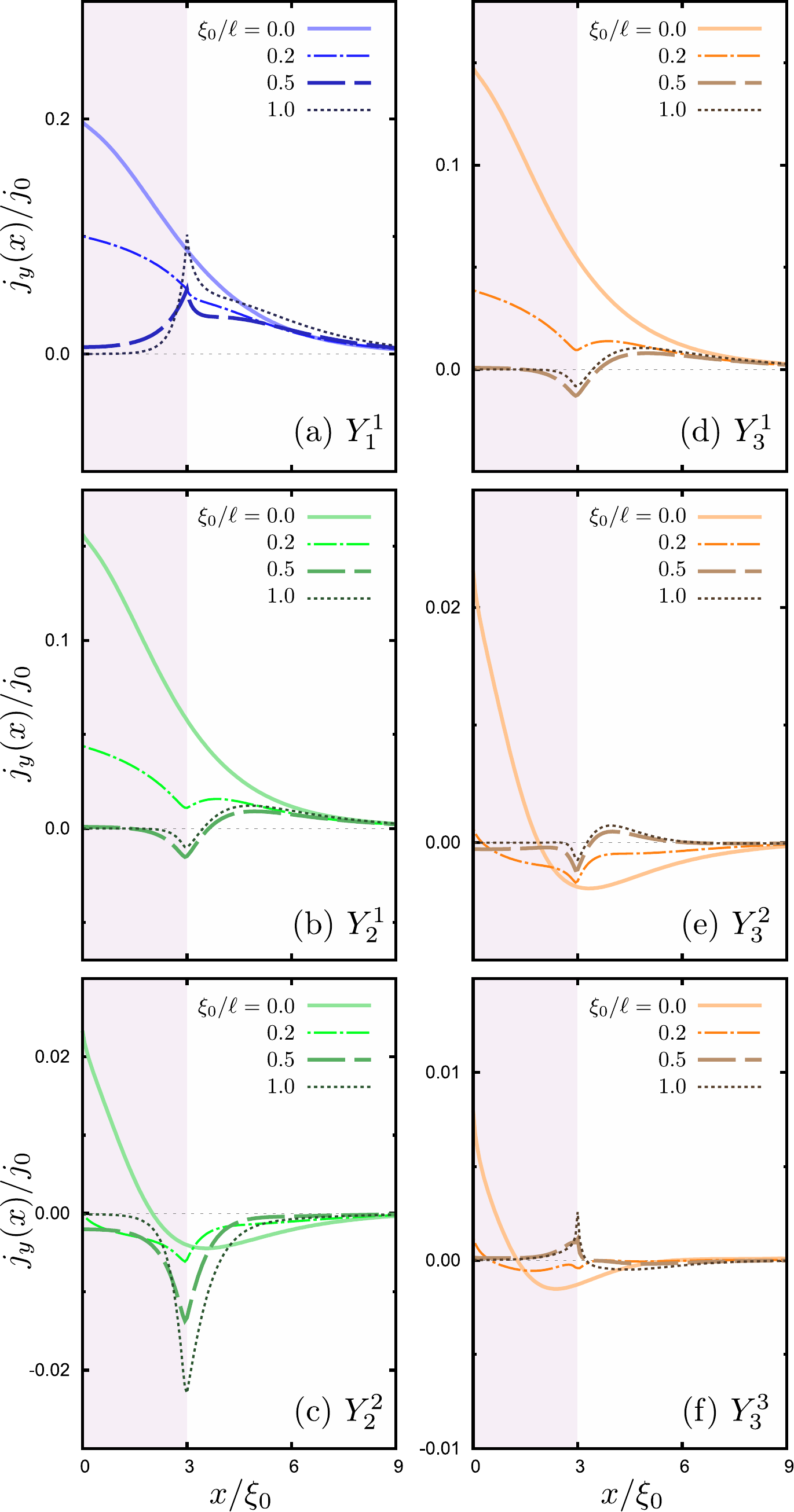}
	\caption{Roughness dependence of the total current density in each
	chiral superconductors.  The strength of the surface roughness is
	set to $\xi_0/\ell = 0$, $0.2$, $0.5$, and $1.0$. The other
	parameters are set to the same values as used in Fig.~\ref{fig:cur}. 
	When the subdominant $s$-wave pairs are present, the chiral current
	flows along the internal interface as in (a,c,f). On the other hand, the chiral
	current without the $s$-wave pairs is smeared out by the roughness
  as in (b,d,e).  
	}
  \label{fig:cur_rhodep}
\end{figure}

\begin{figure}[tb]
	\includegraphics[width=0.46\textwidth]{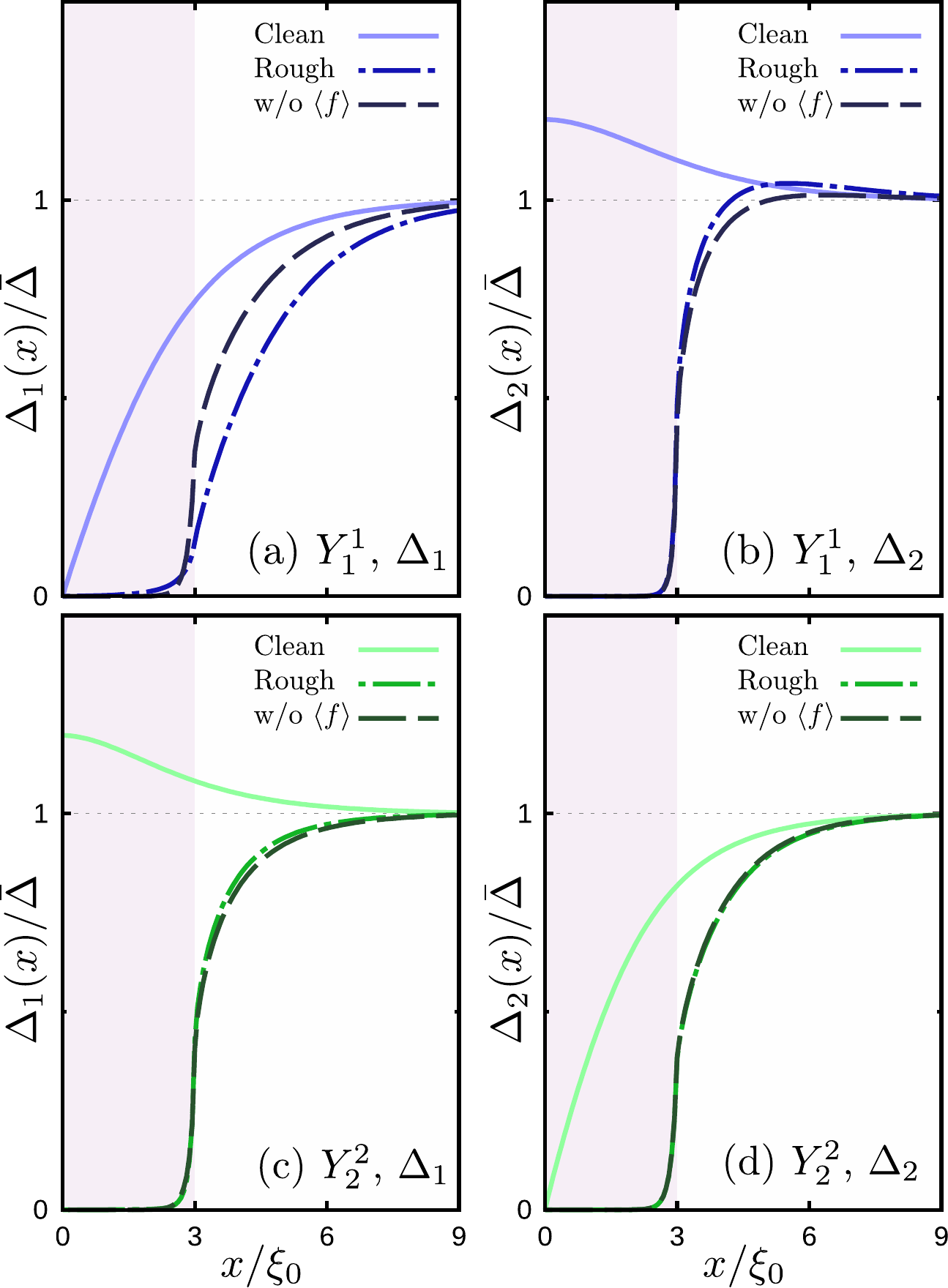}
	\caption{Pair potentials for 
		(a,b) $Y_1^1$ and 
		(e,f) $Y_2^2$ superconductors. 
		The first and second components are shown in (a,c) and (b,d). 
		The surface roughness significantly suppresses the pair potential
		near the surface. 
		The amplitudes of the suppression depend on the presence of
		$\langle f \rangle$. 
	}
  \label{fig:del}
\end{figure}

To examine the effect of the $s$-wave Cooper pairs $\langle f \rangle$, we calculate
the chiral surface currents under the condition $\langle f \rangle =
0$ in Eq.~\ref{eq:ep}. 
This condition is not realistic but clarifies the effect of the
$s$-wave subdominant pairs $\langle f \rangle$.  The results are shown in Fig.~\ref{fig:cur}
with the broken lines. 
Even in the $Y_1^1$ SC, the chiral current is significantly suppressed
if there ware no $s$-wave Cooper pair $\langle f \rangle$. The chiral currents in the
$Y_2^2$ and $Y_3^3$ SCs are also significantly suppressed compared with the
full calculation with $\langle f \rangle$. From these 
 simulation, we 
conclude that the robust chiral current under surface roughness 
is supported by the $s$-wave
subdominant Cooper pairs induced by the disorder. 

The roughness dependences of the chiral surface current are shown in 
Figs.~\ref{fig:cur_rhodep}, where the parameters on the surface
roughness are set to $w=3\xi_0$ and $\xi_0/\ell = 0.0$, $0.2$, $0.5$,
and $1.0$. 
In the $Y_1^1$ and $Y_2^2$ case, the peak of the current
density moves from the surface to the interface with increasing the
roughness [Fig.~\ref{fig:cur_rhodep}(a,c)]. The chiral current in those
SCs would be observed regardless of the surface quality of the sample. 
In the $Y_3^3$ case, the amplitude of the chiral current remains
finite even in the rough surface. However, the amplitude is small even
in the clean limit. It is not clear if the chiral surface current in
the $Y_3^3$ SC can be observed in experiments. 
When the $s$-wave pairs $\langle f \rangle$ are absent or sufficiently
small,
the chiral surface current is easily destroyed even by the week
roughness (e.g., $\ell > \xi_0$) [Fig.~\ref{fig:cur_rhodep}(b,d,e)].
The observed chiral current would be
much smaller than that estimated in the clean limit. Therefore, one
needs to fabricated a sample with a specular surface to observe the
chiral current in these SCs.

The existence of the $s$-wave subdominant SC affects also the pair
potentials. The self-consistent pair potentials are shown in
Fig.~\ref{fig:del}, where the strength of the disorder is set to
$\xi_0/\ell = 5$. The results for the $Y_1^1$ and $Y_2^2$ 
are shown in Figs.~\ref{fig:del}(a,b) and \ref{fig:del}(c,d)
\footnote{The pair potentials for the other chiral SCs are
qualitatively the same as those of the $Y_2^2$ SC because the
subdominant $s$-wave amplitude is not large enough to affect the pair
potentials}. 
In the clean limit, the emergence of 
the Andreev bound states suppress one of
the pair potential depending on the momentum dependence
[See, Figs.~\ref{fig:del}(a,c)] \cite{Matsumoto_JPSJ_1999}. 
One of the pair potentials that changes its sign during the
quasiparticle reflection at the surface (e.g., $p_x$-wave component of
$Y_1^1$ SC) is responsible to form the
ABSs.  
Correspondingly, the other component of the pair
potential slightly enhanced at the surface
[Figs.~\ref{fig:del}(b,c)]. When the surface is rough, both of the
components are significantly suppressed. The chiral
superconductivity is realized by anisotropic Cooper pairs which are
fragile against impurity scatterings (e.g., Anderson's theorem
\cite{Anderson}). 
The roughness dependence of the pair potentials are discussed in the
Appendix.

The calculated pair potentials without the $s$-wave pairs $\langle f \rangle$ are also
shown in Fig.~\ref{fig:del}. In the $Y_1^1$ case, $\Delta_1$ with $\langle f
\rangle$ is smaller than that without $\langle f \rangle$. This
behavior means the $s$-wave Cooper pairs induces the effective pair
potential in the disordered region and makes the internal interface an
effective boundary. The other component of
the pair potentials is also affected by the $s$-wave Cooper pairs. 
The similar behavior can be seen in the $Y_2^2$ case. However,
the effect is much smaller than the $Y_1^1$ case because the amplitude of
the $s$-wave pairs is much smaller [see Fig.~\ref{fig:pair}(b)]. We do
not show the results for $Y_2^1$,  $Y_3^1$,  $Y_3^2$, and  $Y_3^3$
because the $s$-wave subdominant pairs are absent or sufficiently small. The
details of the spatial profiles of the pair potentials are discussed
in the Appendix.

\section{Normal-metal surface}
\begin{figure}[tb]
	\includegraphics[width=0.46\textwidth]{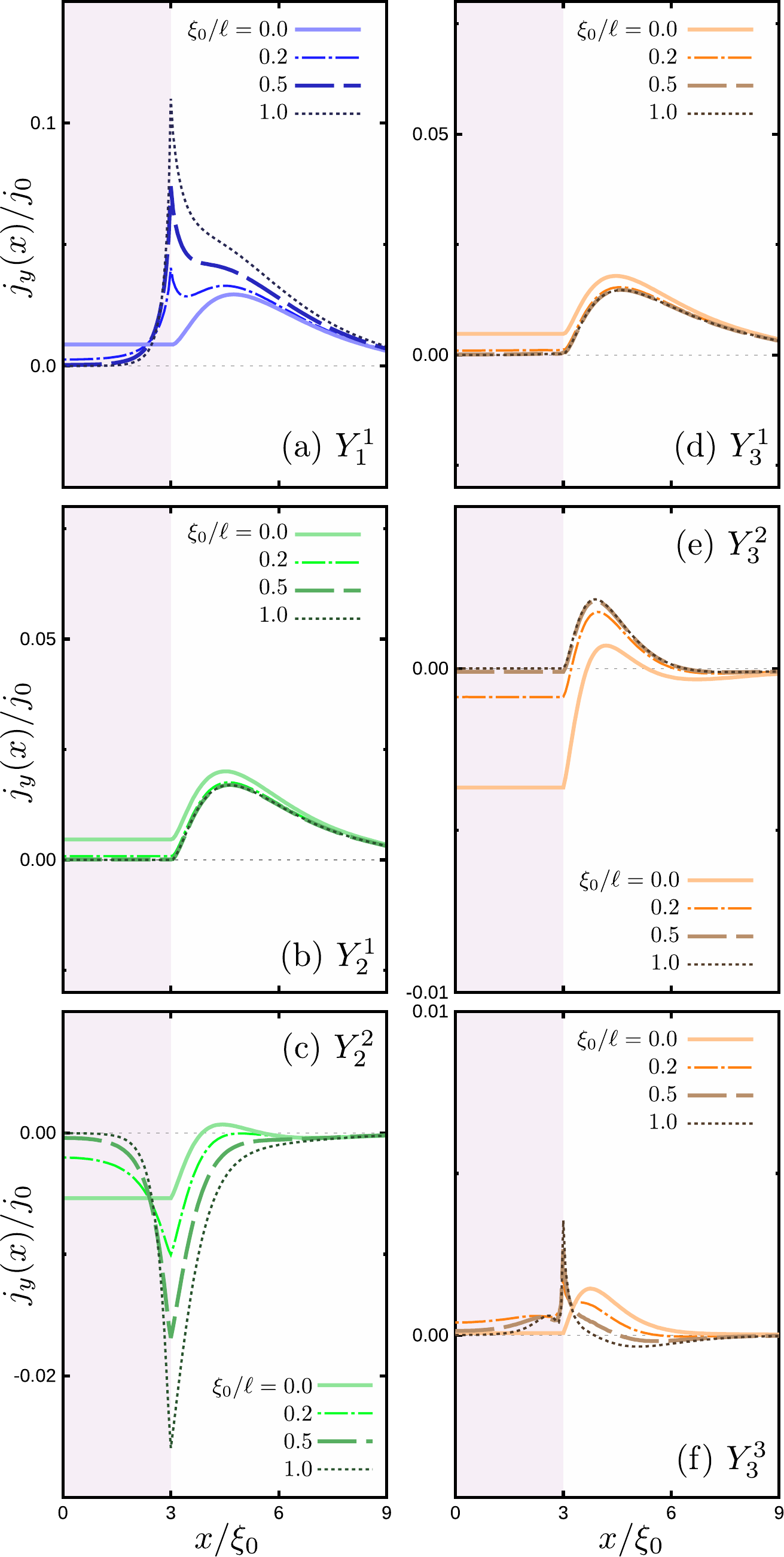}
	\caption{Current density of each chiral SC with \textit{normal}
	surfaces. The results are plotted in the same manner as in Fig.~3. 
	When the $s$-wave pairs have a large amplitude, the chiral current
	can survive as in (a,c,f). }
  \label{fig:cur_norm}
\end{figure}

The surface roughness affects the chiral surface current also through
the modifying the pair potential near the surface. To examine the
effect through the pair potential, we have introduced the normal-metal
surface [Fig.~\ref{fig:Sche}(b)] where the coupling constant is zero
(i.e., $\Delta_{1(2)}=0$). A superconductor with a thin dirty
normal-metal surface corresponds to the so-called Ovchinnikov model 
\cite{Ovchinnikov}.
The chiral surface currents with a thin normal-metal surface are shown in
Fig.~\ref{fig:cur_norm}, where the results are plotted in the same
manner as in Fig.~\ref{fig:cur_rhodep}. 
{We have confirmed the pair potentials do not depend strongly on
the roughness parameter $\xi_0/\ell$ (see the Appendix). } 
The normal metal with $w=3\xi_0$ is 
indicated by the shaded regions. 

When the surface is in a clean metallic state, the chiral surface current
is small and would not be able to be measured in experiments
regardless of the symmetry of the SC as shown in
Fig.~\ref{fig:cur_norm}.\cite{Lederer_PRB_14, Bakurskiy_17}
When the surface is a clean normal metal, $j_y$ does not depend on $x$ in the
normal-metal because of $\Delta=0$. 
Compared with that in a 2D model \cite{Bakurskiy_17}, the surface effect in a
3D model is more prominent because there are more channels with the
low injection angles. Those quasiparticles travel in the disordered
region longer than those with high injection angles. 

When the surface normal metal is disordered, the larger chiral
current flows in the $Y_1^1$ SCs compared with the results with 
the clean-metal surface (i.e., $\xi_0/\ell=0$). In the $Y_2^2$ and
$Y_3^3$ SCs, the total current does not seem
to change with increasing the roughness even though the spatial
dependence changes drastically. 
The current density in the $Y_2^1$, $Y_3^1$, and $Y_3^2$ SCs are
simply suppressed with increasing the roughness. 
The difference comes from whether the $s$-wave pairs are
present in the surface normal metal. 
As happened in the rough-surface SCs [Figs.~\ref{fig:cur}, 
\ref{fig:pair}, and \ref{fig:cur_rhodep}], the $s$-wave subdominant pairs
cause the effective superconductivity and the chiral current flows
along the internal interface [Fig.~\ref{fig:cur_norm}(a,c,f)]. 
When there is no (or significantly small) $s$-wave pairing in the
normal metal, no effective superconductivity is expected, leading that
the chiral current does not arise along the internal interface
[Fig.~\ref{fig:cur_norm}(b,d,e)]. Note that a finite current flows
inside the SC (e.g., $Y_2^1$ and $Y_3^1$) and similar profiles appear in the rough-surface model
[Fig.~\ref{fig:cur}]. This current emerges because of the spatial
gradient of the pair potential. This current density is, however, much
smaller than those estimated in the clean limit. 

Comparing the results with a rough surface and those with a
normal-metal surface, we see that the existence of the $s$-wave
subdominant pairs are essential for the observable chiral current. The
suppression of the pair potential is less important than the presence
of the $s$-wave pairs. Namely, we can judge the robustness of the
chiral current by whether the $s$-wave pairs are induced as a
subdominant Cooper pairs at a surface.

\section{Conclusion}

We have studied the effects of the surface roughness on the
spontaneous chiral surface current of the general chiral SC utilizing the
quasiclassical Eilenberger theory. We have considered a
three-dimensional chiral SC with a rough or a dirty-normal-metallic
surface, where the pair potential is generally described by the
spherical harmonics $Y_l^m$ with a finite magnetic quantum number ($m 
\neq 0$). 

From the self-consistent solutions, the spontaneous current in $Y_1^1$
 and $Y_2^2$ (i.e., $p_x + i p_y$- and $d_{x^2-y^2}+id_{xy}$-wave) SCs
 are sufficiently large to detect in experiments even under the
strong surface roughness. That in the $Y_3^3$ 
[i.e., $f_{x(x^2-3y^2)} + i f_{y(3x^2-y^2)}$-wave] SC survives under the 
strong roughness but might be too small to be observed. 
The surface chiral currents in $Y_2^1$, $Y_2^1$, and $Y_2^1$ SCs are easily destroyed even by 
weak surface roughness (i.e., $\ell > \xi_0$). 

Comparing the calculated current densities with and without the
subdominant $s$-wave Cooper pairs induced by the roughness, we have
concluded that the robust chiral current is supported by the $s$-wave
Cooper pairs. The $s$-wave Cooper pairs act as an effective pair
potential in the disordered region. The current density in this case
flows along the effective interface between the disordered and clean region that
can be regarded as an interface between an effective SC and the chiral
SC.

\begin{acknowledgments}
  We are grateful to Y.~Asano for the fruitful discussions. 
  S.-I.~S. is supported by 
  JSPS Postdoctoral Fellowship for Overseas Researchers
  and a Grant-in-Aid for JSPS Fellows
  (JSPS KAKENHI Grant No. JP19J02005),  
  and would like to thank the University of Twente for hospitality.
\end{acknowledgments}


\appendix

\begin{figure}[tb]
	\includegraphics[width=0.46\textwidth]{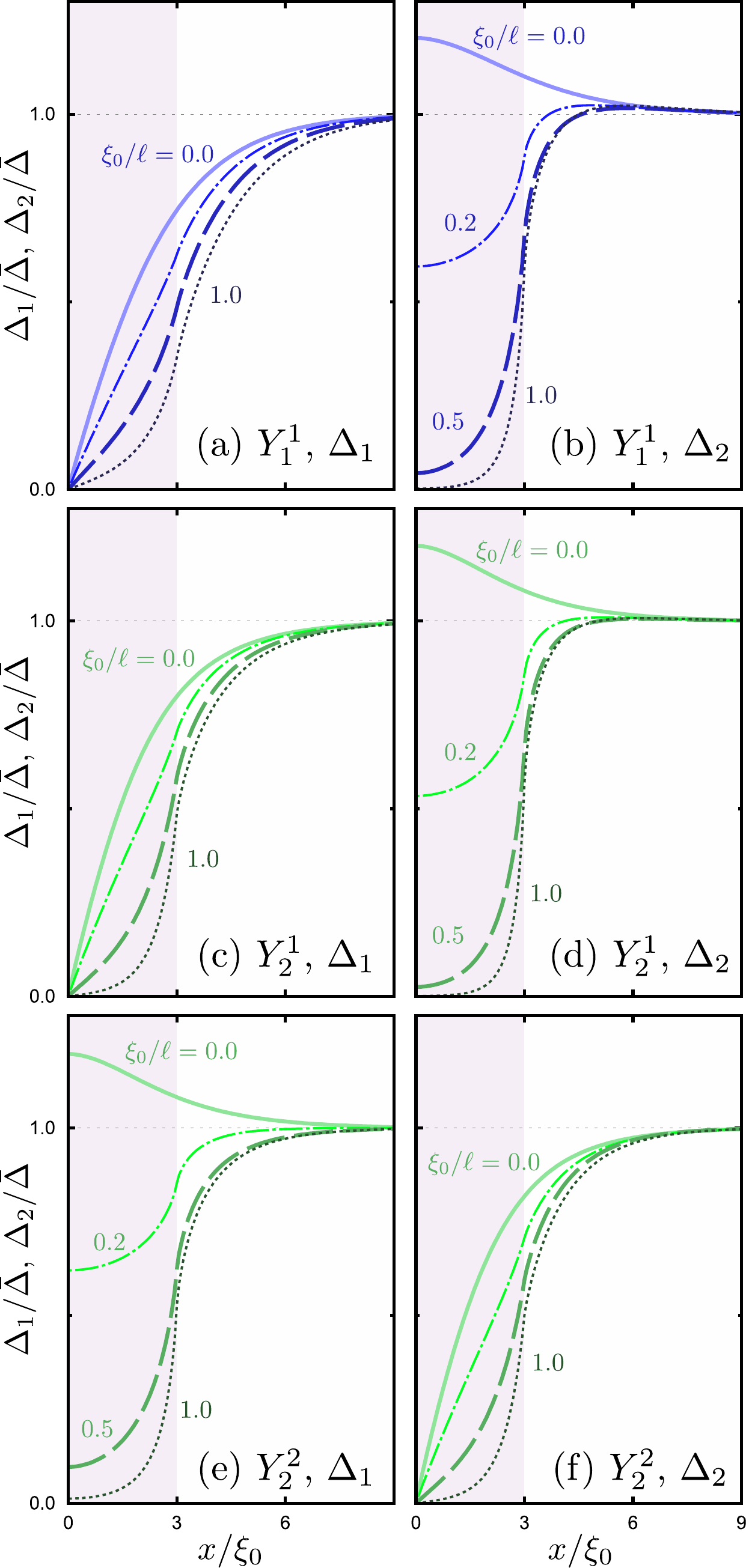}
	\caption{Roughness dependence of the pair potentials for 
		(a,b) $Y_1^1$, 
		(c,d) $Y_2^1$, and 
		(e,f) $Y_2^2$. 
		The first and second components are shown in (a,c,e) and (b,d,f). 
		The surface-roughness parameters are set to $w=3\xi_0$, $\xi_0
		/\ell = 0.0$, $0.2$, $0.5$, and $1.0$. 
		The surface roughness significantly suppresses the pair potential. 
	}
  \label{fig:del_rhodep}
\end{figure}
\begin{figure}[tb]
	\includegraphics[width=0.48\textwidth]{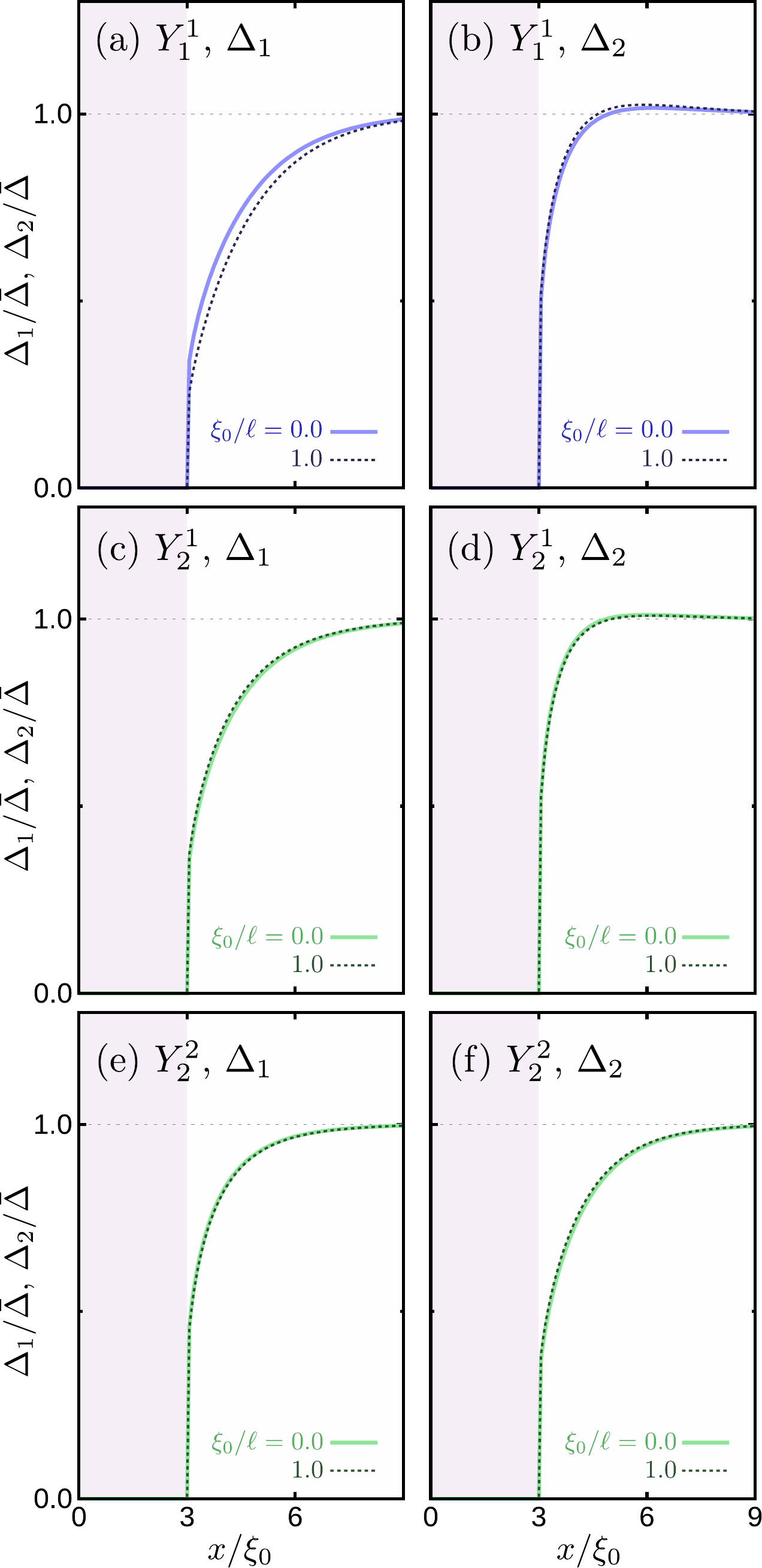}
	\caption{
  Roughness dependence of the pair potentials with the thin dirty
	normal metal. The pair potentials are assumed 
		(a,b) $Y_1^1$, 
		(c,d) $Y_2^1$, and 
		(e,f) $Y_2^2$. 
		The first and second components are shown in (a,c,e) and (b,d,f). 
		The surface-roughness parameters are set to $w=3\xi_0$, $\xi_0
		/\ell = 0.0$, and $1.0$. 
	The results are plotted in
  the same manner as in Fig.~\ref{fig:del_rhodep}. 
	}
  \label{fig:del_norm}
\end{figure}

\section{Effect of the surface roughness on Pair potentials}

The spatial dependences of the pair potential in SCs with rough
surfaces are shown in Fig.~\ref{fig:del_rhodep}, where the strength of
the roughness is set to $\xi_0/\ell = 0$, $0.2$, $0.5$, and $1.0$. 
The pair potentials are suppressed by the surface roughness through
the random scatterings. The suppression in the pair potentials becomes
more significant with increasing the roughness strength. The chiral
superconductivity realizes with anisotropic Cooper pairing. The random
quasiparticle scatterings kill such pairings with anisotropic momentum
dependence.
As a consequence, the pair potential decreases in the disordered
region. The spatial dependences of the pair potentials of the $Y_l^m$
SCs with $l=3$ (not shown) are similar to those in the $Y_2^1$ SC
because the amplitude of the $s$-wave pairs are zero or sufficiently
small.

The spatial dependences of the pair potential in SCs with
dirty-normal-metal surfaces are shown in Fig.~\ref{fig:del_norm},
where the strength of
the roughness is set to $\xi_0/\ell = 0$ or $1.0$. 
Differing from the results with a rough surface, the pair potentials
do not strongly depend on the strength of the roughness. 
The spatial dependences of the pair potentials of the $Y_l^m$ SCs with
$l=3$ (not shown) are similar to those in the $Y_2^1$ SC because the
amplitude of the $s$-wave pairs are zero or sufficiently small.


\begin{thebibliography}{2}%
\makeatletter
\providecommand \@ifxundefined [1]{%
 \@ifx{#1\undefined}
}%
\providecommand \@ifnum [1]{%
 \ifnum #1\expandafter \@firstoftwo
 \else \expandafter \@secondoftwo
 \fi
}%
\providecommand \@ifx [1]{%
 \ifx #1\expandafter \@firstoftwo
 \else \expandafter \@secondoftwo
 \fi
}%
\providecommand \natexlab [1]{#1}%
\providecommand \enquote  [1]{``#1''}%
\providecommand \bibnamefont  [1]{#1}%
\providecommand \bibfnamefont [1]{#1}%
\providecommand \citenamefont [1]{#1}%
\providecommand \href@noop [0]{\@secondoftwo}%
\providecommand \href [0]{\begingroup \@sanitize@url \@href}%
\providecommand \@href[1]{\@@startlink{#1}\@@href}%
\providecommand \@@href[1]{\endgroup#1\@@endlink}%
\providecommand \@sanitize@url [0]{\catcode `\\12\catcode `\$12\catcode
  `\&12\catcode `\#12\catcode `\^12\catcode `\_12\catcode `\%12\relax}%
\providecommand \@@startlink[1]{}%
\providecommand \@@endlink[0]{}%
\providecommand \url  [0]{\begingroup\@sanitize@url \@url }%
\providecommand \@url [1]{\endgroup\@href {#1}{\urlprefix }}%
\providecommand \urlprefix  [0]{URL }%
\providecommand \Eprint [0]{\href }%
\providecommand \doibase [0]{http://dx.doi.org/}%
\providecommand \selectlanguage [0]{\@gobble}%
\providecommand \bibinfo  [0]{\@secondoftwo}%
\providecommand \bibfield  [0]{\@secondoftwo}%
\providecommand \translation [1]{[#1]}%
\providecommand \BibitemOpen [0]{}%
\providecommand \bibitemStop [0]{}%
\providecommand \bibitemNoStop [0]{.\EOS\space}%
\providecommand \EOS [0]{\spacefactor3000\relax}%
\providecommand \BibitemShut  [1]{\csname bibitem#1\endcsname}%
\let\auto@bib@innerbib\@empty
\bibitem [{Note1()}]{Note1}%
  \BibitemOpen
  \bibinfo {note} {The quasiclassical approximation allows us to extract the
  essential spatial profile (i.e., coherence-length order) from the Green's
  function by ignoring the rapid oscillation with the Fermi-wavelength
  order.}\BibitemShut {Stop}%
\bibitem [{Note2()}]{Note2}%
  \BibitemOpen
  \bibinfo {note} {The pair potentials for the other chiral SCs are
  qualitatively the same as those of the $Y_2^2$ SC because the subdominant
  $s$-wave amplitude is not large enough to affect the pair
  potentials}\BibitemShut {NoStop}%
\end{thebibliography}%


\begin{thebibliography}{99}


\bibitem{Kita_JPSJ_1998} T. Kita, 
	Angular Momentum of Anisotropic Superfluids at Finite Temperatures,
	\href{https://doi.org/10.1143/jpsj.67.216}{
	J. Phys. Soc. Jpn. \textbf{67}, 216 (1998)}. 

\bibitem{Matsumoto_JPSJ_1999}M.~Matsumotoand and M.~Sigrist, 
Quasiparticle states near the surface and the domain wall in a $p_x \pm ip_y$-wave superconductor,
{J. Phys. Soc. Jpn. \textbf{68} 3 (1999)}. 

\bibitem{Furusaki_PRB_2001}A.~Furusaki, M.~Matsumotoand and M.~Sigrist,
Spontaneous Hall effect in a chiral $p$-wave superconductor,
\href{https://link.aps.org/doi/10.1103/PhysRevB.64.054514}
{Phys. Rev. B \textbf{64}, 054514 (2001)}. 
%
\bibitem{Stone_04} M.~Stone and R.~Roy, 
Edge modes, edge currents, and gauge invariance in $p_x + i p_y$ superfluids and superconductors, 
\href{https://doi.org/10.1103/PhysRevB.69.184511}
{Phys. Rev. B \textbf{69}, 184511 (2004)}.
%
\bibitem{Nagato_11} Y.~Nagato, S.~Higashitani, and K.~Nagai, 
Subgap in the Edge States of Two-Dimensional Chiral Superconductor with Rough Surface
\href{https://doi.org/10.1143/JPSJ.80.113706}{J. Phys. Soc. Jpn.  \textbf{80}, 113706 (2011)}. 
%

\bibitem{Sauls_PRB_2011} J. A. Sauls, 
Surface states, edge currents, and the angular momentum of chiral
$p$-wave superfluids, 
\href{https://doi.org/10.1103/PhysRevB.84.214509}{
Phys. Rev. B \textbf{84}, 214509 (2011)}.

\bibitem{Bakurskiy_14}S. V. Bakurskiy, A. A. Golubov, M. Yu. Kupriyanov, K. Yada, and Y. Tanaka, 
Anomalous surface states at interfaces in $p$-wave superconductors, 
\href{https://doi.org/10.1103/PhysRevB.90.064513}{Phys. Rev. B \textbf{90}, 064513 (2014)}.

\bibitem{Sigrist_PRB_2014} A. Bouhon and M. Sigrist, 
	Current inversion at the edges of a chiral p-wave superconductor, 
	\href{https://doi.org/10.1103/PhysRevB.90.220511}{
	Phys. Rev. B \textbf{90}, 220511(R) (2014)}.

\bibitem{Lederer_PRB_14}
	S. Lederer, W. Huang, E. Taylor, S. Raghu, and C. Kallin, 
Suppression of spontaneous currents in Sr2RuO4 by surface disorder
\href{https://doi.org/10.1103/PhysRevB.90.134521}{
Phys. Rev. B \textbf{90}, 134521 (2014)}.

\bibitem{Bakurskiy_17} S. V. Bakurskiy, N. V. Klenov, I. I. Soloviev, M. Yu. Kupriyanov, and A. A. Golubov, 
Observability of surface currents in p-wave
superconductors, 
\href{https://doi.org/10.1088/1361-6668/aa5f3d}
{Supercond. Sci. Technol. \textbf{30}, 044005 (2017)}.


\bibitem{Braunecker_PRL_2005} B. Braunecker, P. A. Lee, and Z. Wang, 
Edge Currents in Superconductors with a Broken Time-Reversal Symmetry
\href{https://doi.org/10.1103/PhysRevLett.95.017004}{
Phys. Rev. Lett. \textbf{95}, 017004 (2005)}. 

\bibitem{Huang_PRB_2014} W. Huang, E. Taylor, and C. Kallin, 
Vanishing edge currents in non-p-wave topological chiral superconductors, 
\href{https://doi.org/10.1103/PhysRevB.90.224519}{
Phys. Rev. B \textbf{90}, 224519 (2014)}.

\bibitem{Tada_PRL_2015} Y. Tada, W. Nie, and M. Oshikawa, 
Orbital Angular Momentum and Spectral Flow in Two-Dimensional Chiral Superfluids,
\href{https://doi.org/10.1103/PhysRevLett.114.195301}{
Phys. Rev. Lett. \textbf{114}, 195301 (2015). }

\bibitem{suzuki_16}S.-I.~Suzuki and Y.~Asano, 
Spontaneous edge current in a small chiral superconductor with a rough surface, 
\href{https://doi.org/10.1103/PhysRevB.94.155302}{Phys. Rev. B
\textbf{94}, 155302 (2016)}. 

\bibitem{Wang_PRB_2018} X. Wang, Z. Wang, and C. Kallin, 
	Spontaneous edge current in higher chirality superconductors, 
	\href{https://doi.org/10.1103/PhysRevB.98.094501}{
	Phys. Rev. B \textbf{98}, 094501 (2018)}.


\bibitem{Higashitani_JPSJ_20} E.~Sugiyama and S.~Higashitani, 
Surface Bound States and Spontaneous Edge Currents in Chiral
Superconductors: Effect of Spatially Varying Order Parameter, 
\href{https://doi.org/10.7566/JPSJ.89.034706}
{J. Phys. Soc. Jpn. \textbf{89}, 034706 (2020)}.

\bibitem{Nie_PRB_20}
	W. Nie, W. Huang, and H. Yao, 
Edge current and orbital angular momentum of chiral superfluids
revisited, 
\href{https://doi.org/10.1103/PhysRevB.102.054502}{
Phys. Rev. B \textbf{102}, 054502 (2020)}. 
 
\bibitem{SIS_PRR_2022}
	S.-I. Suzuki, S. Ikegaya, and A. A. Golubov, 
	Destruction of surface states of $(d_{zx}+id_{yz})$-wave
	superconductor by surface roughness: Application to Sr$_2$RuO$_4$,
	\href{https://doi.org/10.1103/PhysRevResearch.4.L042020}{
	Phys. Rev. Research \textbf{4}, L042020 (2022)}. 

%
\bibitem{moler_05}P. G. Bj\"ornsson, Y. Maeno, M. E. Huber, and K. A. Moler,
Scanning magnetic imaging of Sr$_2$RuO$_4$,
\href{https://journals.aps.org/prb/abstract/10.1103/PhysRevB.72.012504}{Phys. Rev. B \textbf{72}, 012504 (2005)}.
%
\bibitem{nelson_07}J. R. Kirtley, C. Kallin, C. W. Hicks, E. -A. Kim, Y. Liu, K. A. Moler, Y. Maeno, and K. D. Nelson,
Upper limit on spontaneous supercurrents in Sr$_2$RuO$_4$,
\href{https://journals.aps.org/prb/abstract/10.1103/PhysRevB.76.014526}{Phys. Rev. B \textbf{76}, 014526 (2007)}.
%


\bibitem{kobayashi_15} S. Kobayashi, Y. Tanaka, and M. Sato,
Fragile surface zero-energy flat bands in three-dimensional chiral superconductors,
\href{https://journals.aps.org/prb/abstract/10.1103/PhysRevB.92.214514}{Phys. Rev. B \textbf{92}, 214514 (2015)}.
%
\bibitem{tamura_17} S. Tamura, S. Kobayashi, L. Bo, and Y. Tanaka,
Theory of surface Andreev bound states and tunneling spectroscopy in three-dimensional chiral superconductors,
\href{https://journals.aps.org/prb/abstract/10.1103/PhysRevB.95.104511}{Phys. Rev. B \textbf{95}, 104511 (2017)}.
%
\bibitem{suzuki_20} S.-I. Suzuki, M. Sato, and Y. Tanaka,
Identifying possible pairing states in Sr$_2$RuO$_4$ by tunneling spectroscopy,
\href{https://journals.aps.org/prb/abstract/10.1103/PhysRevB.101.054505}{Phys. Rev. B \textbf{101}, 054505 (2020)}.
%



\bibitem{Tanaka_PRL_2003}
A. Tanaka and X. Hu, 
Possible Spin Triplet Superconductivity in Na$_x$CoO$_2$$\cdot y$H$_2$O, 
\href{https://doi.org/10.1103/PhysRevLett.91.257006}{
Phys. Rev. Lett. \textbf{91}, 257006 (2003)}. 

\bibitem{Baskaran_PRL_2003} G. Baskaran, 
Electronic Model for CoO$_2$ Layer Based Systems: Chiral Resonating
Valence Bond Metal and Superconductivity,
\href{https://doi.org/10.1103/PhysRevLett.91.097003}{
Phys. Rev. Lett. \textbf{91}, 097003 (2003)}.


\bibitem{takada2003superconductivity}
K. Takada, H. Sakurai, E. Takayama-Muromachi, F. Izumi, R. A. Dilanian and T. Sasaki, 
Superconductivity in two-dimensional CoO$_2$ layers,
\href{https://doi.org/10.1038/nature01450}{
Nature 422, \textbf{53} (2003)}. 


\bibitem{Ogata_JPSJ_2003}
M. Ogata, 
Superconducting States in Frustrating $t-J$ Model:
A Model Connecting High-$T_c$ Cuprates, Organic Conductors and
Na$_x$CoO$_2$, 
\href{https://doi.org/10.1143/JPSJ.72.1839}{
J. Phys. Soc. Jpn. \textbf{72}, 1839 (2003)}.

\bibitem{Kiesel_PRL}
M. L. Kiesel, C. Platt, W. Hanke, and R. Thomale
Model Evidence of an Anisotropic Chiral d+id-Wave Pairing State for
the Water-Intercalated Na$_x$CoO$_2$$\cdot y$H$_2$O Superconductor, 
\href{https://doi.org/10.1103/PhysRevLett.111.097001}{
Phys. Rev. Lett. \textbf{111}, 097001 (2013)}.



\bibitem{Kiesel_PRB_12} M. L. Kiesel, C. Platt, W. Hanke, D. A. Abanin, and R. Thomale, 
Competing many-body instabilities and unconventional superconductivity in graphene, 
\href{https://doi.org/10.1103/PhysRevB.86.020507}{
Phys. Rev. B \textbf{86}, 020507(R) (2012)}. 

\bibitem{BS_PRL_12} A. M. Black-Schaffer, 
	Edge Properties and Majorana Fermions in the
	Proposed Chiral $d$-Wave Superconducting State of Doped Graphene, 
	\href{https://doi.org/10.1103/PhysRevLett.109.197001}{
	Phys. Rev. Lett. \textbf{109}, 197001 (2012)}. 

\bibitem{Nand_NatPhys_12} R. Nandkishore, L. S. Levitov, and A. V. Chubukov, 
Chiral superconductivity from repulsive
interactions in doped graphene, 
\href{http://www.nature.com/doifinder/10.1038/nphys2208}{
Nat. Phys. \textbf{8}, 158 (2012)}. 

\bibitem{BS_JPhys_14} A. M. Black-Schaffer and C. Honerkamp, 
	Chiral $d$-wave superconductivity in doped graphene, 
	\href{http://dx.doi.org/10.1088/0953-8984/26/42/423201}{
J. Phys.: Condens. Matter \textbf{26}, 423201 (2014)}. 



\bibitem{Biswas_PRB_13}
P. K. Biswas, H. Luetkens, T. Neupert, T. Stürzer, C. Baines, G.
Pascua, A. P. Schnyder, M. H. Fischer, J. Goryo, M. R. Lees, H.
Maeter, F. Brückner, H.-H. Klauss, M. Nicklas, P. J. Baker, A. D.
Hillier, M. Sigrist, A. Amato, and D. Johrendt, 
Evidence for superconductivity with broken time-reversal symmetry in
locally noncentrosymmetric SrPtAs, 
\href{http://dx.doi.org/10.1103/PhysRevB.87.180503}{
Phys. Rev. B \textbf{87}, 180503(R) (2013)}. 


\bibitem{Fischer_PRB_14}
	M. H. Fischer, T. Neupert, C. Platt, A. P. Schnyder,
	W. Hanke, J. Goryo, R. Thomale, and M. Sigrist, 
	Chiral $d$-wave superconductivity in SrPtAs, 
	\href{https://doi.org/10.1103/PhysRevB.89.020509}{
	Phys. Rev. B \textbf{89}, 020509(R) (2014)}. 

\bibitem{Landaeta_PRB_16}
J. F. Landaeta, S. V. Taylor, I. Bonalde, C. Rojas, Y. Nishikubo, K.
Kudo, and M. Nohara, 
High-resolution magnetic penetration depth and inhomogeneities in
locally noncentrosymmetric SrPtAs, 
\href{http://dx.doi.org/10.1103/PhysRevB.93.064504}{
Phys. Rev. B \textbf{93}, 064504 (2016)}. 

\bibitem{Goryo_PRB_17}
	J. Goryo, Y. Imai, W. B. Rui, M. Sigrist, and A. P. Schnyder, 
Surface magnetism in a chiral $d$-wave superconductor with hexagonal symmetry, 
\href{https://doi.org/10.1103/PhysRevB.96.140502}{
Phys. Rev. B \textbf{96}, 140502(R) (2017)}. 

\bibitem{Ueki_PRB_19}
H. Ueki, R. Tamura, and J. Goryo
Possibility of chiral $d$-wave state in the hexagonal pnictide
superconductor SrPtAs, 
\href{https://doi.org/10.1103/PhysRevB.99.144510}{
Phys. Rev. B \textbf{99}, 144510 (2019)}. 



\bibitem{Kasahara_PRL_07}
Y. Kasahara, T. Iwasawa, H. Shishido, T. Shibauchi, K. Behnia, Y.
Haga, T. D. Matsuda, Y. Onuki, M. Sigrist, and Y. Matsuda, 
Exotic Superconducting Properties in the Electron-Hole-Compensated
Heavy-Fermion “Semimetal” URu$_2$Si$_2$, 
\href{http://dx.doi.org/10.1103/PhysRevLett.99.116402}{
Phys. Rev. Lett. \textbf{99}, 116402 (2007)}. 

\bibitem{Kasahara_NJP_09}
Y. Kasahara, H. Shishido, T. Shibauchi, Y. Haga, T. D. Matsuda, Y.
Onuki, and Y. Matsuda, 
Superconducting gap structure of heavy-Fermion compound URu$_2$Si$_2$ determined by angle-resolved thermal conductivity, 
\href{https://dx.doi.org/10.1088/1367-2630/11/5/055061}{
New J. Phys. \textbf{11}, 055061 (2009)}. 

\bibitem{Kittaka_JPSJ_16}
S. Kittaka, Y. Shimizu, T. Sakakibara, Y. Haga, E. Yamamoto, Y.
$\bar{O}$nuki, Y. Tsutsumi, T. Nomoto, H. Ikeda, and K. Machida, 
Evidence for Chiral d-Wave Superconductivity in URu$_2$Si$_2$ from the
Field-Angle Variation of Its Specific Heat, 
\href{https://doi.org/10.7566/JPSJ.85.033704}{
J. Phys. Soc. Jpn. \textbf{85}, 033704 (2016)}. 



\bibitem{maeno_94} Y. Maeno, H. Hashimoto, K. Yoshida, S. Nishizaki, T. Fujita, J. G. Bednorz, and F. Lichtenberg,
Superconductivity in a layered perovskite without copper,
\href{https://www.nature.com/articles/372532a0}{Nature (London) \textbf{372}, 532 (1994)}.
\bibitem{maeno_03} A. P. Mackenzie and Y. Maeno,
The superconductivity of Sr$_2$RuO$_4$ and the physics of spin-triplet pairing,
\href{https://journals.aps.org/rmp/abstract/10.1103/RevModPhys.75.657}{Rev. Mod. Phys. \textbf{75}, 657 (2003)}.
\bibitem{Pustogow_Nature_19} A. Pustogow, Y. Luo, A. Chronister, Y.-S. Su, D. A. Sokolov, F. Jerzembeck, A. P. Mackenzie, C. W. Hicks, N. Kikugawa, S. Raghu, E. D. Bauer, and S. E. Brown,
Constraints on the superconducting order parameter in Sr$_2$RuO$_4$ from oxygen-$17$ nuclear magnetic resonance,
\href{https://www.nature.com/articles/s41586-019-1596-2}{Nature \textbf{574}, 72-75 (2019)}.

\bibitem{Agterberg_PRR_2020} H. G. Suh, H. Menke, P. M. R. Brydon, C. Timm, A.
	Ramires, and D. F.  Agterberg, Stabilizing even-parity chiral
	superconductivity in Sr$_2$RuO$_4$, 
\href{https://doi.org/10.1103/PhysRevResearch.2.032023}{
Phys. Rev. Research \textbf{2}, 032023(R) (2020)}. 

\bibitem{Yuri_PRR_2022}
	Y. Fukaya, T. Hashimoto, M. Sato, Y. Tanaka, and K. Yada, 
Spin susceptibility for orbital-singlet Cooper pair in the three-dimensional Sr2RuO4 superconductor
\href{https://doi.org/10.1103/PhysRevResearch.4.013135}{
Phys. Rev. Research \textbf{4}, 013135 (2022)}.

\bibitem{Grinenko_20} V. Grinenko, S. Ghosh, R. Sarkar, J.-C. Orain, A. Nikitin, M. Elender, D. Das, Z. Guguchia, F. Br\"{u}ckner, M. E. Barber, J. Park, N. Kikugawa, D. A. Sokolov, J. S. Bobowski, T. Miyoshi, Y. Maeno, A. P. Mackenzie, H. Luetkens, C. W. Hicks, and H.-H. Klauss,
Split superconducting and time-reversal symmetry-breaking transitions in Sr$_2$RuO$_4$ under stress,
\href{https://www.nature.com/articles/s41567-021-01182-7}{Nat. Phys. \textbf{17}, 748 (2021)}.
%
\bibitem{Grinenko_21} V. Grinenko, D. Das, R. Gupta, B. Zinkl, N. Kikugawa, Y. Maeno, C. W. Hicks, H.-H. Klauss, M.~Sigrist, and R. Khasanov,
Unsplit superconducting and time reversal symmetry breaking transitions in Sr$_2$RuO$_4$ under hydrostatic pressure and disorder,
\href{https://arxiv.org/abs/2103.03600}{arXiv:2103.03600}.
%

\bibitem{Ikegaya_PRR_21} S. Ikegaya, S.-I. Suzuki, Y. Tanaka, and D. Manske,
Proposal for identifying possible even-parity superconducting states in Sr$_2$RuO$_4$ using planar tunneling spectroscopy,
\href{https://journals.aps.org/prresearch/abstract/10.1103/PhysRevResearch.3.L032062}{Phys. Rev. Research \textbf{3}, L032062 (2021)}.


\bibitem{Fisher_PRL_1989}
R. A. Fisher, S. Kim, B. F. Woodfield, N. E. Phillips, L. Taillefer,
K. Hasselbach, J. Flouquet, A. L. Giorgi, and J. L. Smith, 
Specific heat of UPt$_3$: Evidence for unconventional superconductivity, 
\href{https://doi.org/10.1103/PhysRevLett.62.1411}{
Phys. Rev. Lett. \textbf{62}, 1411 (1989)}. 

\bibitem{Sauls_PRL_1991}
C. H. Choi and J. A. Sauls, 
Identification of odd-parity superconductivity in UPt$_3$ from
paramagnetic effects on the upper critical field, 
\href{https://doi.org/10.1103/PhysRevLett.66.484}{
Phys. Rev. Lett. \textbf{66}, 484 (1991)}.

\bibitem{Sauls_JLTP_1994} J. A. Sauls 
A theory for the superconducting phases of UPt$_3$, 
\href{https://doi.org/10.1007/BF00754932}{
J. Low. Temp. Phys. \textbf{95}, 153 (1994)}. 

\bibitem{Machida_JPSJ_1999}
K. Machida, T. Nishira, and T. Ohmi, 
Orbital Symmetry of a Triplet Pairing in a Heavy Fermion Superconductor UPt$_3$, 
\href{https://doi.org/10.1143/JPSJ.68.3364}{
J. Phys. Soc. Jpn. \textbf{68}, 3364 (1999)}. 

\bibitem{Sauls_PRB_2000}
Matthias J. Graf, S.-K. Yip, and J. A. Sauls, 
Identification of the orbital pairing symmetry in UPt$_3$, 
\href{https://doi.org/10.1103/PhysRevB.62.14393}{
Phys. Rev. B \textbf{62}, 14393 (2000)}.

\bibitem{Joynt_RMP_02}
R. Joynt and L. Taillefer, 
The superconducting phases of UPt3, 
\href{https://doi.org/10.1103/RevModPhys.74.235}{
Rev. Mod. Phys. \textbf{74}, 235 (2002)}. 

\bibitem{Machida_PRL_12}
Y. Machida, A. Itoh, Y. So, K. Izawa, Y. Haga, E. Yamamoto, N. Kimura,
Y. Onuki, Y. Tsutsumi, and K. Machida, 
Twofold Spontaneous Symmetry Breaking in the Heavy-Fermion
Superconductor UPt$_3$, 
\href{https://doi.org/10.1103/PhysRevLett.108.157002}{
Phys. Rev. Lett. \textbf{108}, 157002 (2012)}. 

\bibitem{Tsutsumi_JPSJ_12}
Y. Tsutsumi, K. Machida, T. Ohmi, and M. Ozaki, 
A Spin Triplet Superconductor UPt$_3$, 
\href{https://doi.org/10.1143/JPSJ.81.074717}{
J. Phys. Soc. Jpn. \textbf{81}, 074717 (2012)}.

\bibitem{Izawa_JPSJ_14}
K. Izawa, Y. Machida, A. Itoh, Y. So, K. Ota, Y. Haga, E. Yamamoto, 
N. Kimura, Y. Onuki, Y. Tsutsumi, and K. Machida, 
Pairing Symmetry of UPt$_3$ Probed by Thermal Transport Tensors, 
\href{https://doi.org/10.7566/JPSJ.83.061013}{
J. Phys. Soc. Jpn. \textbf{83}, 061013 (2014)}. 

\bibitem{Lambert}
F. Lambert, A. Akbari., P. Thalmeier, and I. Eremin, 
Surface State Tunneling Signatures in the Two-Component Superconductor
UPt$_3$, 
\href{https://doi.org/10.1103/PhysRevLett.118.087004}{
Phys. Rev. Lett. \textbf{118}, 087004 (2017)}. 



\bibitem{Yamada_JPSJ_96}
	K. Yamada, Y. Nagato, S. Higashitani, and K. Nagai, 
	Rough Surface Effects on d-Wave Superconductors, 
	\href{
	https://doi.org/10.1143/JPSJ.65.1540}{
	J. Phys. Soc. Jpn. \textbf{65}, 1540 (1996)}. 

\bibitem{Nagai_JPSJ_08}
	K. Nagai, Y. Nagato, M. Yamamoto, and S. Higashitani, 
	Surface Bound States in Superfluid $^3$He, 
\href{https://doi.org/10.1143/JPSJ.77.111003}{J. Phys. Soc. Jpn.
\textbf{77}, 111003 (2008)}. 

\bibitem{SIS_PRB_15}
	S.-I. Suzuki and Y. Asano, 
Effects of surface roughness on the paramagnetic response of small
unconventional superconductors, 
\href{https://doi.org/10.1103/PhysRevB.91.214510}{
Phys. Rev. B \textbf{91}, 214510 (2015)}. 

\bibitem{Higashitani_JPSJ_15}
S. Higashitani and N. Miyawaki, 
Phase Transition to a Time-Reversal Symmetry-Breaking State in
$d$-Wave Superconducting Films with Rough Surfaces, 
\href{https://doi.org/10.7566/JPSJ.84.033708}{
J. Phys. Soc. Jpn. \textbf{84}, 033708 (2015)}. 


%
\bibitem{Eilenberger}G.~Eilenberger, 
Transformation of Gorkov’s equation for type II superconductors into transport-like equations,
\href{https://doi.org/10.1007/BF01379803}{Z. Physik \textbf{214}, 195–213 (1968)}. 
%
\bibitem{Schopohl_PRB_95}N.~Schopohl and K.~Maki, 
Quasiparticle spectrum around a vortex line in a $d$-wave superconductor,
\href{https://link.aps.org/doi/10.1103/PhysRevB.52.490}
{Phys. Rev. B \textbf{52}, 490 (1995)}. 
%
\bibitem{Eschrig_PRB_00}M.~Eschrig, 
Distribution functions in nonequilibrium theory of superconductivity
and Andreev spectroscopy in unconventional superconductors, 
\href{https://link.aps.org/doi/10.1103/PhysRevB.61.9061}
{Phys. Rev. B \textbf{61}, {9061--9076} (2000)}. 
%
\bibitem{Eschrig_PRB_09} M.~Eschrig, 
{Scattering problem in nonequilibrium quasiclassical theory of metals and superconductors: General boundary conditions and applications},
\href{https://link.aps.org/doi/10.1103/PhysRevB.80.134511}
{Phys. Rev. B \textbf{80}, {134511} (2009)}. 
%
\bibitem{Anderson} P. W. Anderson, 
Theory of dirty superconductors, 
\href{https://doi.org/10.1016/0022-3697(59)90036-8}
{J. Phys. Chem. Solids. \textbf{11}: 26–30 (1959)}. 

\bibitem{Ovchinnikov}
Y. N. Ovchinnikov Zh. Eksp. Teor. Fiz. 56 1590 (1969) 
[Sov. Phys. JETP \textbf{29}, 853 (1969)]. 

\bibitem{Golubov_Ovchi98}
A.~A.~Golubov and M.~Yu.~Kupriyanov, 
Anomalous proximity effect in $d$-wave superconductors, 
\href{https://doi.org/10.1134/1.567717}
{JETP Lett. \textbf{67}, 501 (1998)}. 

\bibitem{Golubov_Ovchi99}
A.~A.~Golubov and M.~Yu.~Kupriyanov, 
Surface electronic scattering in d-wave superconductors, 
\href{https://doi.org/10.1134/1.568015}
{JETP Letters \textbf{69}, 262 (1999)}. 


\end{thebibliography}
\end{document}